\documentclass[12pt, margin=1in]{article}

\usepackage[margin=1in]{geometry}

\usepackage{amsmath}
\usepackage{booktabs}
\usepackage{graphicx}
\usepackage{subcaption}
\usepackage{microtype}

\usepackage{titlesec}
\titleformat{\section}
  {\sffamily\bfseries\large}
  {\thesection}{1em}{}

\titleformat{\subsection}
  {\sffamily\bfseries\normalsize}
  {\thesubsection}{1em}{}

\titleformat{\subsubsection}
  {\sffamily\bfseries\normalsize}
  {\thesubsubsection}{1em}{}

\usepackage{caption}

\usepackage{lmodern}
\usepackage{placeins}

\usepackage{algorithm}
\usepackage{algpseudocode}

\usepackage[table,dvipsnames]{xcolor}
\usepackage[most]{tcolorbox}
\tcbuselibrary{skins, breakable}
\tcbset{
  commonstyle/.style={
    enhanced,
    breakable,
    boxrule=0.6pt,
    arc=6pt,
    left=10pt,
    right=10pt,
    top=8pt,
    bottom=8pt,
    colback=black!2,
    colframe=black!25,
    fonttitle=\bfseries\small,
    fontupper=\small,
  }
}

\usepackage{xurl}

\usepackage[numbers,sort&compress]{natbib}
\bibpunct{[}{]}{,}{a}{}{,}

\usepackage{float}

\usepackage{hyperref}

\hypersetup{
    hidelinks,
    hypertexnames=false,
    breaklinks=true
}
\definecolor{citegreen}{RGB}{0,110,70}
\hypersetup{
    colorlinks=true,
    citecolor=citegreen
}
  \definecolor{claudebg}{RGB}{250,235,225}
  \definecolor{gptbg}{RGB}{225,235,250}
  \definecolor{owbg}{RGB}{235,240,230}

  \title{Agent trajectories as programs: fingerprinting and programming coding-agent behavior}
  \author{Hamidah Oderinwale\thanks{Email: hamidah.oderinwale@mail.mcgill.ca, Work done with support from Taste Labs}}
  \date{November 2026}

\begin{document}
  \maketitle

\begin{abstract}
Benchmark scores tell you what an agent got right; they do not tell you how it got there. In this work, we introduce methods for
  comparing agents procedurally in different contexts, where the model, tasks, and approaches vary. We compare ten agents and
  find that they are identifiable by their behavioral habits, which we define as \emph{fingerprints}: a probe over these
  procedural signatures attributes an unseen trajectory to the correct agent at 85.7\% accuracy.
  We develop procedural representations for agent problem-solving procedures with an emergent vocabulary induction technique which
  is meant to be maximally compressive to avoid surface-level variation while being expressive enough to unveil the quirks of the
  models' patterns. We apply our framework to the software engineering evaluation dataset SWE-Bench to study the structural
  distinctness of agent trajectories and find that behavior is most similar between models from similar release periods and those
  that are distilled from one another (e.g., a distilled student model and its teacher have a Jensen-Shannon divergence of 0.25, about half the distance between other model pairs). As more models saturate evaluations, we believe that it will be important to probe
  model behavior along more holistic dimensions than success rates alone. We introduce \texttt{ProcGrep}, a library for auditing and
  evaluating agents for how they approach tasks at a procedural level given their traces in a top-down fashion. One application of ProcGrep is searching over past events. We show that ProcGrep outperforms LLMs on episodic search in both efficiency and accuracy, offering a deterministic, programmable approach to searching over traces. We believe this
  work has a range of applications to help developers work with and program coding agents, such as task-aware model routing, agent
  monitoring, and finer-grained cost analysis.\footnote{\texttt{ProcGrep} is available at
  \url{https://github.com/hamidahoderinwale/procgrep}.}
\end{abstract}

  %  \section{Introduction}  As programming becomes more high-level, program analysis needs to adapt beyond the syntax
  %  level to
  % accommodate agents who can not only be optimized by the code they write but the actions they take. Improving agents
  % will depend on procedural
  % programmability and minimizing the gap between human intent an agent action. Additionally, it will involve assessing
  % trajectories and holistic evaluations beyond binary outcomes \citep{haupt2025position}.

  \section{Background}
  As agent-style programming becomes more accessible, engineers will spend less time writing code directly and more time
  shaping how their agents frame and solve problems in the form of \emph{scaffolds}---the wrappers around models that
  determine which tools they can employ, how they acquire context, and how they react to feedback. That said, traditional
  program analysis methods are too static to support this reliably and at scale
  \citep{jimenez2024swebenchlanguagemodelsresolve, yang2024sweagentagentcomputerinterfacesenable,
  chen2021evaluatinglargelanguagemodels, austin2021programsynthesislargelanguage}. In this work, we introduce a suite of
  procedural programming tools and analyses on agent traces from software engineering evaluations to address this problem.
  We motivate this work by showing how variation in agent behavior affects outcomes. While traces can be interpreted in
  isolation, there is much to gain from studying them in aggregate---tapping into collective wisdom
  \citep{surowiecki2004wisdom, Alizadeh_2015}---yet there is a lack of tools for doing so, as meaningful comparison calls
  for shared representations.

  While chain-of-thought has proven itself (eliciting rationale from the agent) useful for post-hoc rationalizing---it
  cannot be taken as grounded accounts of action \citep{wei2023chainofthoughtpromptingelicitsreasoning,
  ross2025modelingstudentlearning38, guan2025monitoringmonitorability}. Given this, in this work we explore representations
  from traces to turn them into structured objects and motivate the need for doing so, and we employ our suite of
  analytical tools to study the different trajectories that agents take outside controlled settings (in an area of work we
  call \emph{behavioral fingerprinting}). The goal is to expand the evaluation of agents beyond whether they simply pass or
  fail a task by venturing into the analysis of their procedural habits or ``style'' \citep{bisztray2025iknowllmwrote}.
  Additionally, we propose a number of interfaces and measurement frameworks to make these kinds of studies more
  accessible. This is our motivation for building ProcGrep which is a framework for setting up agent studies and studying
  their procedural variation.

  Previous work has explored solution variation at scale for MOOCs (massively online open courses) to observe behavior with
  clustering \citep{glassman2015overcode}. Platforms like LMArena compare agent outputs by evaluating preferred outputs in a
  pairwise fashion \citep{chiang2024chatbotarenaopenplatform}. Prior work has investigated using LLMs to generate plans
  from formal specifications~\citep{silver2024generalized}; with procedural representations it is possible to do the
  inverse, where queries can be written as procedural specs to `grep'
  traces.~\citep{yin2017syntacticneuralmodelgeneralpurpose}. Observational studies of LLM chats have been performed at scale
  to understand trends in model use \citep{tamkin2024clioprivacypreservinginsightsrealworld, chatterji2025how}. Frameworks
  have also been introduced for configuring models as agents to perform multi-step tasks with their own feedback loops
  given a goal specified upfront with `discovered' end states \citep{yao2023reactsynergizingreasoningacting}. And past work
  has also shown that coding style can be derived from source code using ASTs and decision trees in the form of random
  forest classifiers \citep{code_stylometry}.

  Existing frameworks have been built for code alone in siloed environments where context is missing, looking at patches
  \citep{GumTree2014, yin-neubig-2017-syntactic} with ASTs (abstract syntax trees) which generalizes across languages. We
  make use of established parsing methods and extend them to work in multi-modal free-form environments where prompts, for
  example, are written in natural language \citep{cruz2026textastherichestpreferencesignal} demonstrating that a rich set of
  features can be observed and made sense of. Next, given one's ability to audit procedures, there is an avenue for work
  which involves seeing what processes yield the best outputs. A new kind of model routing adapted to preferences could
  blossom by extending on our explorations \citep{hu2024routerbenchbenchmarkmultillmrouting}.

  \section{A theory of procedural understanding}
  Framing agent coding traces as procedures lets us comparatively study and interact with them in ways that are useful for
  benchmarking, evaluating the difficulty of evals themselves, differentiating models by their behavioral fingerprints, and
  performing procedural search. We release a library called ProcGrep to support these use cases. For example, ProcGrep
  supports queries such as the following.

  \begin{tcolorbox}[colback=black!3, colframe=black!30, boxrule=0.4pt, left=10pt, right=10pt, top=8pt, bottom=8pt,
  fontupper=\small]
  ``All Claude-4 trajectories from the past two weeks on Python repositories where \texttt{search\_repo} was followed by
  three or more \texttt{read\_file} calls within the first 8 steps, the agent then made at least one \texttt{edit}, and no
  \texttt{run\_test} occurred before the final \texttt{submit}---on instances with difficulty score above 3.''
  \end{tcolorbox}

  In this work, we use traces for a number of popular software engineering benchmarks and test a number of SOTA model
  families (GPT, Claude, DeepSeek, and Qwen). We use public archives of model traces for SWE-Bench
  (\url{princeton-nlp/SWE-bench-experiments}) and do the same for Agentless logs. Table~\ref{tab:models} lists the full set
  of agents studied: ten agents across four scaffolds (SWE-agent, Agentless, DARS, Moatless) with models from the GPT,
  Claude, DeepSeek, and Qwen families.

  \begin{table}[H]
  \centering\small
  \begin{tabular}{lllr}
  \toprule
  \textbf{Scaffold} & \textbf{Model} & \textbf{Paradigm} & \textbf{$n$} \\
  \midrule
  \rowcolor{claudebg} SWE-agent & Claude-3 Opus        & RLHF dense        & 300 \\
  \rowcolor{claudebg} SWE-agent & Claude-3.5           & RLHF dense        & 289 \\
  \rowcolor{claudebg} SWE-agent & Claude-3.7 (thinking)& Extended thinking & 284 \\
  \rowcolor{claudebg} SWE-agent & Claude-4             & Extended thinking & 288 \\
  \rowcolor{gptbg}    SWE-agent & GPT-4                & RLHF dense        & 300 \\
  \rowcolor{gptbg}    SWE-agent & GPT-4o               & RLHF dense        & 278 \\
  \rowcolor{owbg}     DARS      & DeepSeek-R1          & RL reasoning      & 300 \\
  \rowcolor{owbg}     Agentless & Claude-3.5           & RLHF dense        & 300 \\
  \rowcolor{owbg}     Moatless  & DeepSeek-V3          & MoE pretrain      & 300 \\
  \rowcolor{owbg}     SWE-agent & SWE-agent-LM-32B$^\dagger$ & SFT-distilled (Qwen2.5-32B) & 499 \\
  \bottomrule
  \end{tabular}
  \caption{Models and scaffolds studied with the number of task instances per agent (ten agents).
  $^\dagger$Distilled from the Claude-3.7 teacher (row 3) and analysed in \S\ref{sec:distillation}.}
  \label{tab:models}
  \end{table}

  We define what is the appropriate language to describe and program their behavior. What we call a ``procedural
  fingerprint'' is recovered from what an agent actually did (its trajectory) rather than its components or what it states.
  A fingerprint is a distinguishable feature of an agent from other agents and is consistent over time. Firstly, we
  demonstrate that these fingerprints do in fact exist, and different models are differentiable by their architectural
  types. We find that models are distinguishable by their fingerprints alone at 85.7\% accuracy compared to a random
  baseline of 11.1\% when selecting a model by chance. We attribute this to problem solving \emph{style}
  \citep{bisztray2025iknowllmwrote}. Furthermore, we extend this to deterministic scaffolds such as Agentless and Moatless
  to see if they affect procedural behavior and they also imprint unique fingerprints that are identifiable with near
  perfect accuracy. We performed the comparisons in pairwise settings, to understand distinguishability and we find that
  Claude-4 and SWE-agent-LM-32B are the most identifiable, whereas GPT-4 and Claude-3 Opus are the most confusable (F1
  ~0.20–0.30).

  We show the discriminating action pairs for each agent and demonstrate how much signal for differentiability they offer--in other words, how discriminative they are. We find that deterministic agents have the
  most discriminating pairs. For example, DARS over-uses
  \texttt{search\_repo}$\to$\texttt{create\_file} by $31.6\times$ and Moatless
  \texttt{edit}$\to$\texttt{submit} by $15.7\times$. Interestingly, looking at
  distilled models that are taught via the rollouts of their teacher
  (Claude-3.7 Sonnet), we see some semblance of learned behavior where procedural
  habits are passed down to student (SWE-agent-LM-32B) models.

  \begin{table}[t]
  \centering\small
  \begin{tabular}{llrr}
  \toprule
  \textbf{Agent} & \textbf{Signature transition} & \textbf{Discrim.\ factor ($\times$)} & \textbf{Share} \\
  \midrule
  \rowcolor{owbg}     DARS+R1               & \texttt{search\_repo}$\to$\texttt{create\_file} & 31.6 & 0.022 \\
  \rowcolor{owbg}     Moatless+V3           & \texttt{edit}$\to$\texttt{submit}               & 15.7 & 0.064 \\
  \rowcolor{owbg}     Agentless+Claude-3.5  & \texttt{run\_test}$\to$\texttt{run\_test}       & 12.5 & 0.111 \\
  \rowcolor{claudebg} Claude-3.7-thinking   & \texttt{create\_file}$\to$\texttt{run\_test}    &  9.7 & 0.029 \\
  \rowcolor{owbg}     SWE-agent-LM-32B      & \texttt{create\_file}$\to$\texttt{run\_test}    &  6.0 & 0.023 \\
  \rowcolor{claudebg} Claude-4              & \texttt{read\_file}$\to$\texttt{read\_file}     &  5.0 & 0.603 \\
  \rowcolor{claudebg} Claude-3              & \texttt{create\_file}$\to$\texttt{edit}         &  4.7 & 0.117 \\
  \rowcolor{gptbg}    GPT-4                 & \texttt{create\_file}$\to$\texttt{edit}         &  3.3 & 0.092 \\
  \rowcolor{gptbg}    GPT-4o                & \texttt{run\_test}$\to$\texttt{edit}            &  3.3 & 0.141 \\
  \rowcolor{claudebg} Claude-3.5            & \texttt{run\_test}$\to$\texttt{edit}            &  2.5 & 0.115 \\
  \bottomrule
  \end{tabular}
  \caption{This table ranks the action pairs that distinguish agents. We put the discrimination factor in the third column which shows the
  amount of signal that can be derived from the action pair compared to the random baseline. Share represents the frequency of the transition
  with respect to the total number of action transitions.}
  \label{tab:signature_transitions}
  \end{table}

  \subsection{Inducing action vocabularies}

  Given a set of procedures that we might want to compare, which we can call a procedural space, we need a vocabulary or a
  set of actions that we can use to describe all of them. The contributions of this framework comes from being able to
  define this vocabulary bottom-up instead of with a more hard-coded or classifier based approach. We benchmark our methods
  against prompt-based classifiers, finding that across four judge models, prompt-based classifiers have near-zero agreement across model
  families and struggle to identify compositional trajectories. This affirms our motivations for
  having representations that can be shared across models and instances to allow for comparative analysis (Table~\ref{tab:judge_comparison}).

  This means that there are features of the full vocabulary or alphabet which we can study in itself and can form the basis
  of insights. To define our vocabulary, we need to be able to extract these procedures and label them to understand what
  they are. Each procedure should be composable and collapse into one another because steps are taken in sequence for
  specific tasks and to avoid redundancy---where each procedure of the trajectory should be meaningfully unique or composed
  of unique sub-procedures.   We find that procedural diversity does not correspond with capabilities, where stronger and newer
  extended-thinking
  models do better with fewer procedures (Claude-3.7 = 32, Claude-4 = 35) compared to older Claude-3/3.5 models, which have
  repertoires of 42--44 procedures---which is relatively counter-intuitive.

  \subsection{An information theory of procedures}

  Because of this, we do not rely on a stopping point defined by behavior as in library learning, and instead determine it
  intrinsically. Metrics such as compression ratio, entropy, and vocabulary size characterize a vocabulary but do not on
  their own mark where to stop. We therefore treat the induction of a vocabulary as a similar class of problem to a
  clustering one, and draw on how clustering algorithms are evaluated to form the basis of an emergent stopping criterion
  \citep{rosenberg-hirschberg-2007-v}: the V-measure. We consider a vocabulary stable when it maximizes two measures:
  completeness, meaning the sub-actions of each procedure are correctly aligned with an intent, and homogeneity, meaning a
  procedure contains only actions that belong to it. In our sweep, the V-measure rises, plateaus across vocabulary sizes
  $K=128$--$256$, and then degrades; we take our stopping point at the plateau's peak, $K=192$ (V-measure $0.644$).

  \begin{figure}[H]
      \centering
      \includegraphics[width=0.85\linewidth]{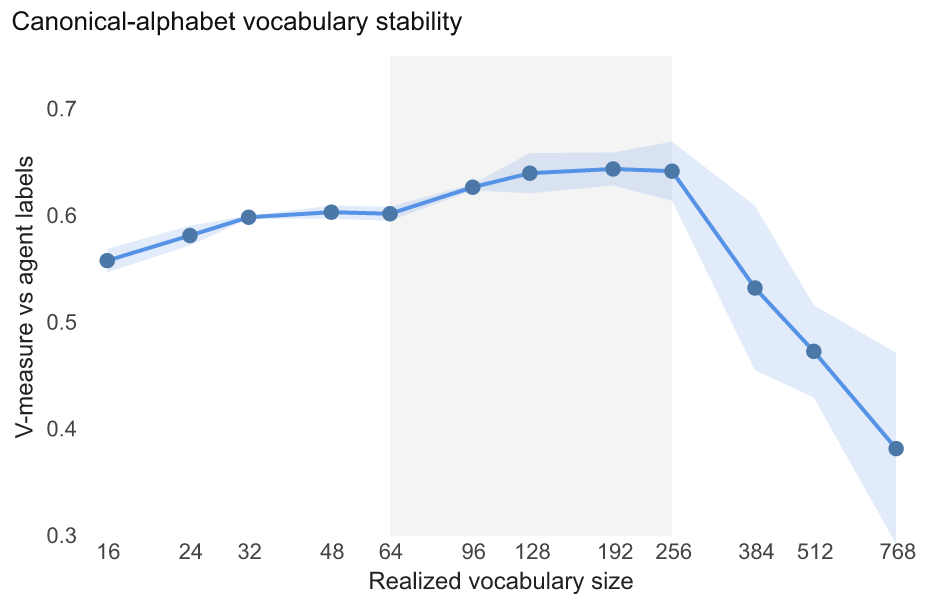}
      \caption{V-measure of BPE vocabulary clustering against agent labels as a function of vocabulary
  size.}
      \label{fig:vmeasure}
  \end{figure}

  BPE or Byte-Pair Encoding is a useful analog for this process and we employ it to induce our vocabulary. While BPE chunks
  natural language tokens by merging those that frequently appear together to compress the data for pre-training LLMs, we
  chunk actions from generated code \citep{sennrich2016neuralmachinetranslationrare}. The idea of learning action
  vocabularies bottom-up is drawn from library learning and program
  synthesis~\citep{ellis2020dreamcodergrowinggeneralizableinterpretable}, where recurring procedural subsequences are
  abstracted from longer programs. While they are typically tied to specific tasks to improve performance, we do so largely
  for the emergent representations they offer. Additionally, BPE is just one algorithm for accomplishing this
  chunking, we evaluate BPE and PrefixSpan along the metrics that we care about in this work, but acknowledge that certain
  methods might favor different goals and find that BPE offers better separability (Figure~\ref{fig:bpe_vs_prefixspan}).

  \begin{figure}[h]
      \centering
      \includegraphics[width=\linewidth]{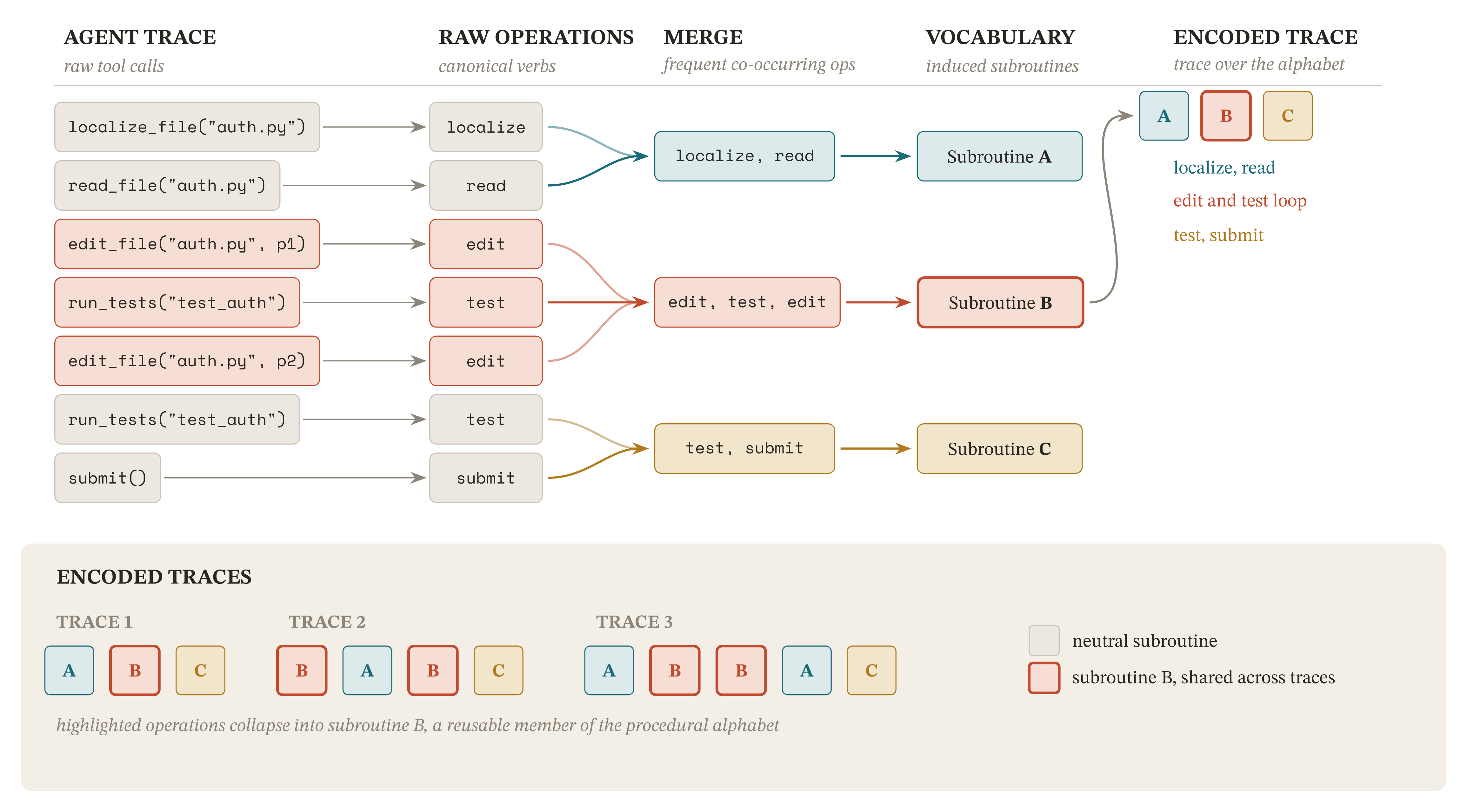}
      \caption{Illustrative figure of the process of vocabulary induction given an agent trace.}
      \label{fig:encoding_pipeline}
  \end{figure}

  \[ V = 2 \cdot \frac{h \cdot c}{h + c} \]
  where homogeneity $h$ and completeness $c$ are defined as
  \[ h = 1 - \frac{H(\text{actions} \mid \text{procedure})}{H(\text{actions})} \qquad c = 1 - \frac{H(\text{procedure} \mid
  \text{actions})}{H(\text{procedure})} \]
  $H(\text{actions} \mid \text{procedure})$ measures the confidence of belonging of an action to a procedure, and
  $H(\text{procedure} \mid \text{actions})$ measures the reverse. A vocabulary is stable when $V$ is maximized. To extract the actions
  themselves, we take code hunks---solution patches---from these evaluation datasets and then parse
  them with an AST that gives a node tree and a syntactical and semantic representation of the code. Given the structural
  representation of the solution patches, we then apply an embedding stage to encode the change in plain language and the
  underlying intent by contextualizing the structured patch with the surrounding code and a behavioral description from the
  model.

We investigate entropy, and use Jensen--Shannon Divergence (JSD) to study how models diverge from the norm. This is
important for detecting out-of-distribution behavior which may be useful in observational settings. Entropy (H), a
measure of uncertainty in a distribution (where a distribution here is the probability over all the possible actions in
our vocabulary), is a property that can be applied to a single agent's distribution of actions, where it is defined as:
$$
H(p_a) = -\sum_{v \in \mathcal{V}} p_a(v) \log_2 p_a(v)
$$
JSD allows us to build on entropy and measure how different two distributions are. We denote the agent distributions $a$ and $b$, where $p_a$ and $p_b$ denote the action distributions of agents $a$ and $b$.

$$
\mathrm{JSD}(p_a, p_b) = H\!\left(\frac{p_a + p_b}{2}\right) - \frac{1}{2}H(p_a)
- \frac{1}{2}H(p_b)
$$

  Across all models, passing behavior includes a higher ratio of browsing to other actions. Extended thinking Claude models
  spend the largest proportion of their time in the shell compared to older models with more even task breakdowns. Models
  in the GPT family have more consistent behavior, but newer ones spend more time editing. Agent harnesses have less task
  diversity, potentially due to their programmed behavior and have more distinct breakdowns, the Agentless agent spends
  a disproportionate amount of time browsing and the Moatless scaffold leads to a large proportion of testing. An
  interesting application of this is forecasting the costs of an agent, as it is not only a product of the cost per tool
  call, but a function of the action types needed to complete it and the procedural tendencies of the model being
  orchestrated.

 \begin{figure}[!t]
    \centering
    \includegraphics[width=0.7\linewidth]{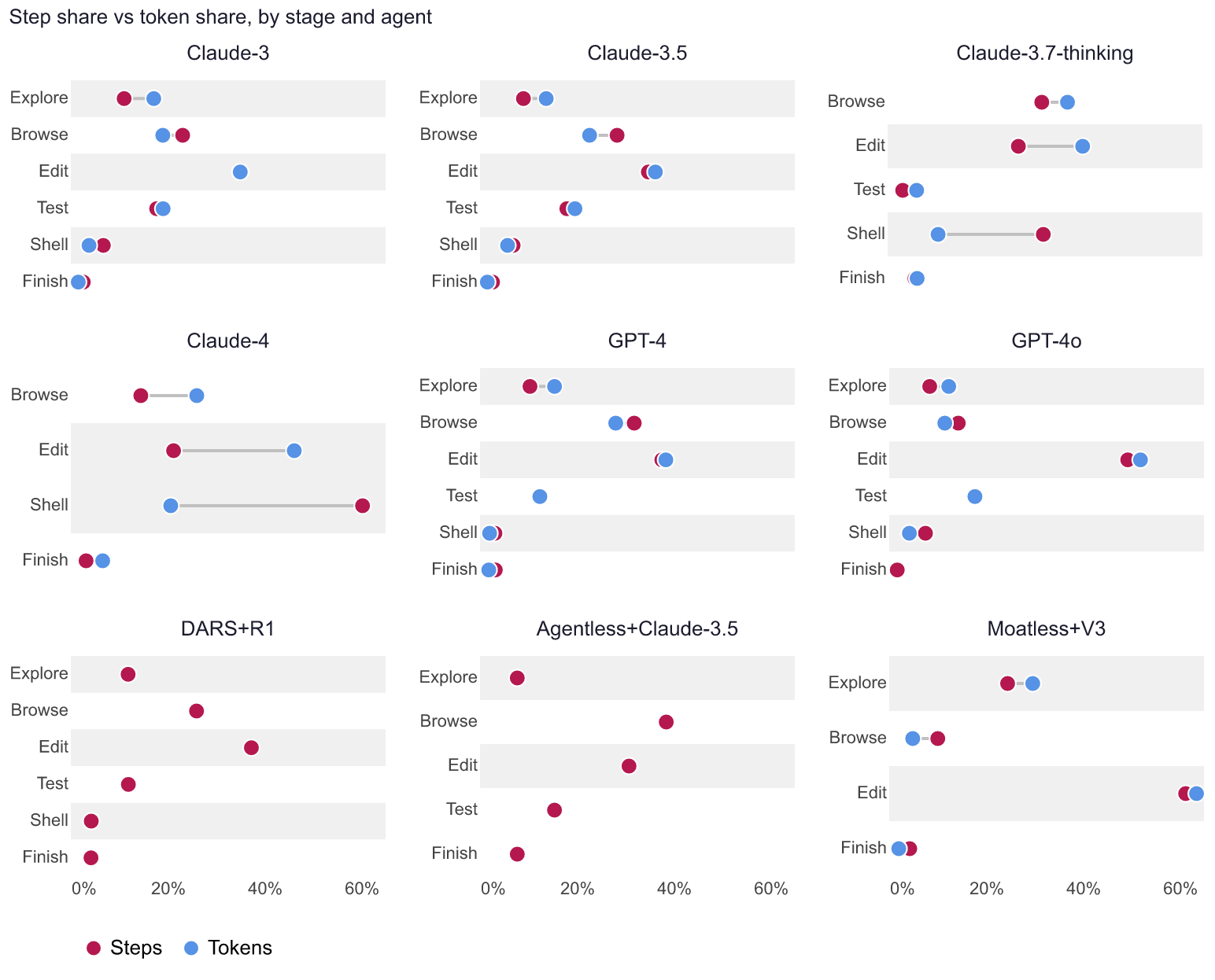}
    \caption{Cost breakdown of agents by action type and tokens.}
    \label{fig:step_token_cost}
\end{figure}

\begin{table*}[!t]
\centering\footnotesize\setlength{\tabcolsep}{4pt}
\resizebox{\linewidth}{!}{%
\begin{tabular}{lllrrrrrrrrr}
\toprule
\textbf{Scaffold} & \textbf{Model} & \textbf{Paradigm} & \textbf{$n$} & \textbf{Resolve} & \textbf{Steps} &
\textbf{Repertoire} & \textbf{Compr.} & \textbf{Entropy} & \textbf{Cost/task} & \textbf{\$/step} & \textbf{Cost/resolved}
\\
 & & & & & \textbf{/traj} & \textbf{@90\%} & \textbf{ratio} & \textbf{(bits)} & & & \\
\midrule
\rowcolor{claudebg} SWE-agent & Claude-3 Opus & RLHF dense & 300 & 11.7\% & 17.1 & 44 & 0.537 & 5.58 & \$3.42 & \$0.200 &
\$29.31 \\
\rowcolor{claudebg} SWE-agent & Claude-3.5 & RLHF dense & 289 & 23.9\% & 32.7 & 42 & 0.559 & 5.56 & \$1.62 & \$0.050 &
\$7.06 \\
\rowcolor{claudebg} SWE-agent & Claude-3.7-thinking & Extended thinking & 284 & 50.7\% & 33.6 & 32 & 0.471 & 5.14 &
\$0.78 & \$0.023 & \$1.53 \\
\rowcolor{claudebg} SWE-agent & Claude-4 & Extended thinking & 288 & 59.0\% & \textbf{64.8} & 35 & 0.417 & 5.26 & \$1.19
& \$0.018 & \$2.02 \\
\midrule
\rowcolor{gptbg} SWE-agent & GPT-4 & RLHF dense & 300 & 18.0\% & 21.4 & 40 & 0.443 & 5.47 & \$2.51 & \$0.117 & \$13.93 \\
\rowcolor{gptbg} SWE-agent & GPT-4o & RLHF dense & 278 & 19.8\% & 39.1 & \textbf{49} & 0.425 & 5.76 & \$2.53 & \$0.065 &
\$13.82 \\
\midrule
\rowcolor{owbg} DARS & DeepSeek-R1 & RL reasoning & 300 & 47.0\% & 24.0 & 19 & \textbf{0.572} & 4.44 & \$2.43$^*$ & --- &
\$5.17$^*$ \\
\rowcolor{owbg} Agentless & Claude-3.5 & RLHF dense & 300 & 40.7\% & 13.0 & \textbf{1} & \textbf{0.077} & \textbf{0.00} &
\$0.50$^*$ & --- & \$1.23$^*$ \\
\rowcolor{owbg} Moatless & DeepSeek-V3 & MoE pretrain & 300 & 30.7\% & 13.1 & 12 & 0.417 & 3.71 & \$0.02 & ${\approx}0$ &
\$0.06 \\
\bottomrule
\end{tabular}%
}
\\[0.3em]
{\footnotesize $^*$Cost estimated. \$/step = cost per inference call. Extended thinking models reduce cost per step by
routing calls through the shell and orchestrating them before invoking the more expensive inference calls (Claude-4:
\$0.018/step vs.\ Claude-3: \$0.200/step).}
\caption{Per-agent procedural fingerprint and cost efficiency on SWE-bench Verified ($n=2{,}639$).}
\label{tab:per-agent-megatable}
\end{table*}

\FloatBarrier

\begin{figure}[H]
    \centering
    \includegraphics[width=0.6\linewidth]{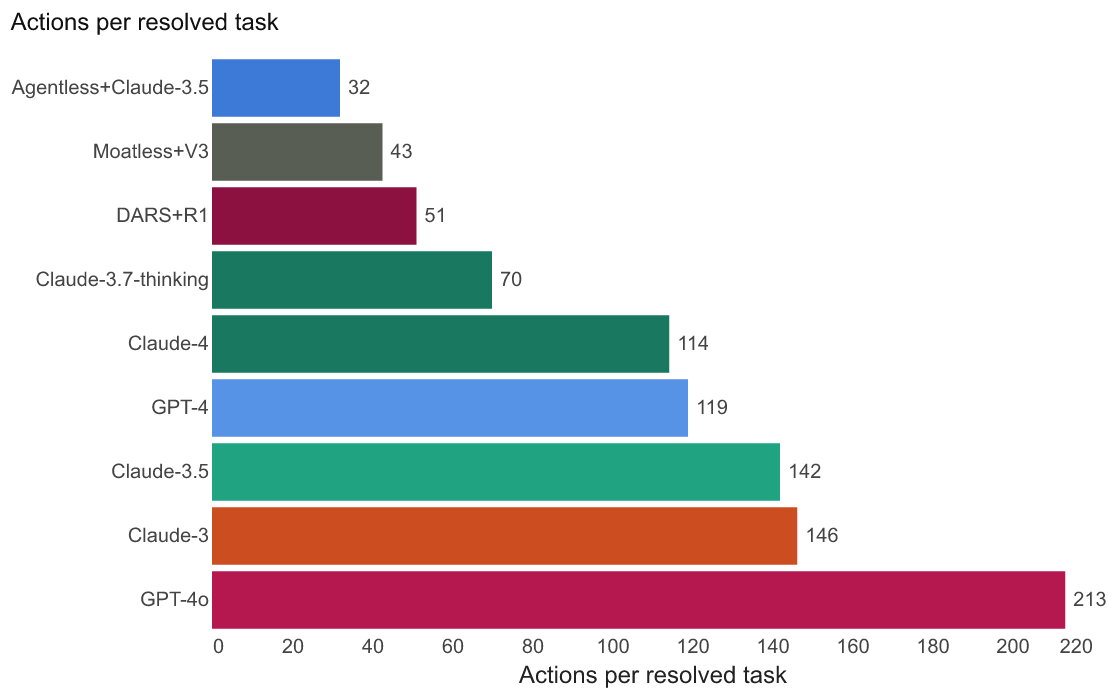}
    \caption{Average number of steps per resolved tasks across agents.}
    \label{fig:steps_resolved}
\end{figure}

\begin{table}[!t]
\centering
\begin{tabular}{lccc}
\toprule
\textbf{Representation} & \textbf{n} & \textbf{F1 (k=1)} & \textbf{Source} \\
\midrule
Structural pattern overlap  & 289 & 0.347 & Deterministic \\
Action sequence distance    & 289 & 0.274 & Deterministic \\
Edit action overlap         & 289 & 0.256 & Deterministic \\
Narrative description       & 289 & 0.177 & Varies by model \\
Agent plan description      & 289 & 0.155 & Varies by model \\
Divergence classifier (LLM) & 87--286 & 0.000--0.363 & $\kappa < 0.05$ (cross-family) \\
\midrule
Random retrieval            & 289 & 0.13--0.24 & Baseline \\
\bottomrule
\end{tabular}
\caption{Comparison of different procedural representations by their ability to
signal whether the agent succeeded or failed, scored with F1 on single
nearest-neighbor prediction. We find that representations derived
directly from the trace beat the random baseline ($0.13$--$0.24$), and that
natural-language accounts barely improve on it.}
\label{tab:representation_comparison}
\end{table}

  \section{Why procedural representations matter}
  Given theoretical grounding, we put foundations to work in a number of ways. In this section we look at top-down analyses
  given agent traces. Firstly, we show that these approaches offer a more grounded representation of process than what can
  be elicited from models in natural language alone. We ask two questions: given what a model says it will do, what does
  it actually do (forward follow-through)? Then, given what a model says it did, what did it actually do (reverse
  follow-through)? Lastly, we show how procedural representations allow for structural search over traces.

  We find that both model families have high reverse follow-through rates, where every action they take is
  mentioned somewhere in their reasoning. For GPT-4 and GPT-4o the reverse follow-through rate is 1.0, and for DARS+R1 and the
  Claude family (Claude-3 / 3.5 / 3.7 / 4) it is high but not complete (0.875 / 0.857 / 0.833 / 0.800 / 0.750, respectively).

  \begin{figure}[H]
      \centering
      \includegraphics[width=0.55\linewidth]{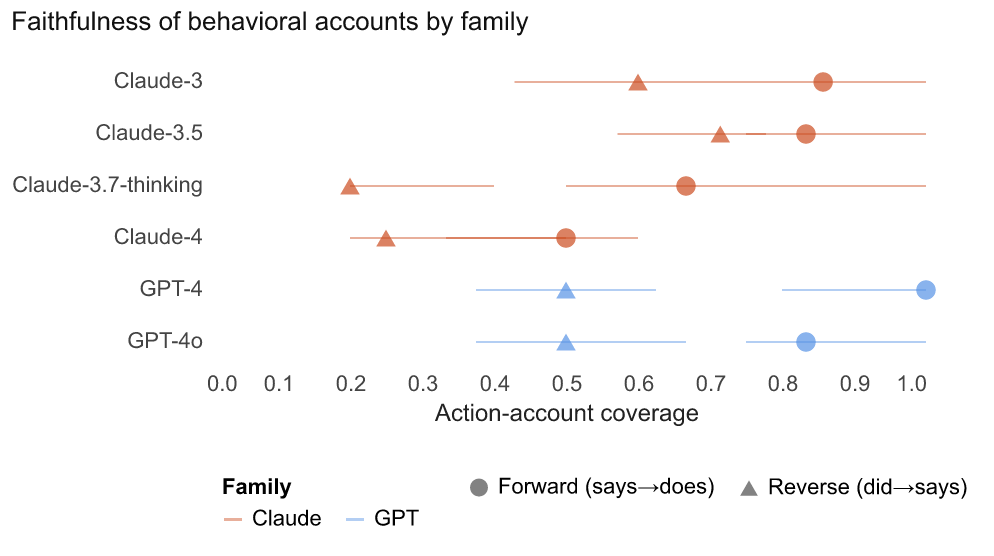}
      \caption{Follow-through for behavioral accounts from the Claude and GPT model families.}
      \label{fig:cot_alignment}
  \end{figure}

  We also assess precision in the behavioral descriptions themselves as a proxy for `groundedness' from a representation. We find that models
  are generally verbose in their
  accounts, which might be a signal of performative chain-of-thought \citep{boppana2026reasoningtheaterdisentanglingmodel}.
  For example, the GPT family are trained with heavy RLHF and instruction-tuning. Whereas the Claude family of models are
  presumably trained to follow more sprawling reasoning patterns before coalescing them into some preferred approach.
  Extended thinking models are the most verbose and have similar behavior to our baseline of shuffled rationales.

  \begin{table}[H]
  \centering\small
  \begin{tabular}{lrrrrr}
  \toprule
  \textbf{Model} & \textbf{n} & \textbf{Verbosity} & \textbf{Precision} & \textbf{Baseline} & \textbf{Recall} \\
  \midrule
  \rowcolor{claudebg} Claude-3            & 300 & 1.07 & 0.833 & 0.678 & 0.857 \\
  \rowcolor{claudebg} Claude-3.5          & 289 & 1.00 & 0.857 & 0.732 & 0.833 \\
  \rowcolor{claudebg} Claude-3.7-thinking & 284 & 1.35 & 0.625 & 0.623 & 0.800 \\
  \rowcolor{claudebg} Claude-4            & 288 & 1.32 & 0.500 & 0.536 & 0.750 \\
  \midrule
  \rowcolor{gptbg}    GPT-4               & 300 & 1.17 & 0.800 & 0.713 & 1.000 \\
  \rowcolor{gptbg}    GPT-4o              & 278 & 1.18 & 0.750 & 0.679 & 1.000 \\
  \midrule
  \rowcolor{owbg} DARS+R1 & 300 & 1.07 & 0.857 & 0.775 & 0.875 \\
  \bottomrule
  \end{tabular}
  \end{table}

 Currently, it is difficult for a developer to search through their agents' history without employing an LLM. The agent first has to surface the code it
  wrote, with context that it may or may not have written to memory. With ProcGrep, search can be deterministic in the same way that writing a SQL query
  is. We compare search with ProcGrep against LLM judges over the same queries, taking the deterministic structural match as ground truth. We find that
  ProcGrep is exact on queries spanning a range of agent actions: context retrieval and the number of files read; conditional events, where an action is
  retrieved based on a preceding event; and missing actions, where an event is retrieved by the absence of an action.

  \begin{table}[H]
    \centering
    \begin{tabular}{lrr}
    \toprule
    Method & Mean $F_1$ & Latency/decision \\
    \midrule
    \rowcolor{owbg} ProcGrep (structural query) & 1.000 & 1.1\,\textmu s \\
    \midrule
    Claude Sonnet 4.6   & 0.278 & 1.71\,s \\
    GPT-4o              & 0.230 & 0.66\,s \\
    GPT-4o-mini         & 0.221 & 0.69\,s \\
    Claude Opus 4.8     & 0.152 & 1.73\,s \\
    Claude 3.5 Haiku    & 0.098 & 1.03\,s \\
    DeepSeek-chat       & 0.093 & 1.51\,s \\
    \bottomrule
    \end{tabular}
    \caption{ProcGrep versus LLM judges.}
    \label{tab:query-vs-llm}
    \end{table}

  \section{Natural studies with procedural representations}
\begin{figure}[H]
  \centering
  \begin{subfigure}[t]{0.48\linewidth}
      \includegraphics[width=\linewidth]{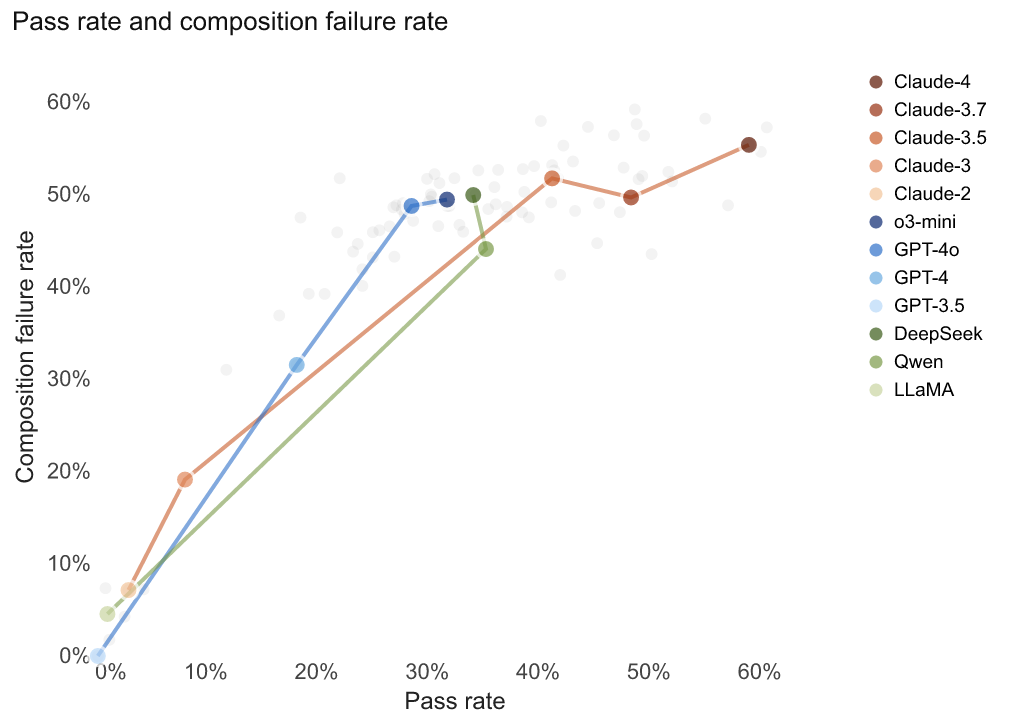}
      \caption{Relationship between compositional problems and pass rate across models.}
      \label{fig:compositional_failure}
  \end{subfigure}
  \hfill
  \begin{subfigure}[t]{0.48\linewidth}
      \includegraphics[width=\linewidth]{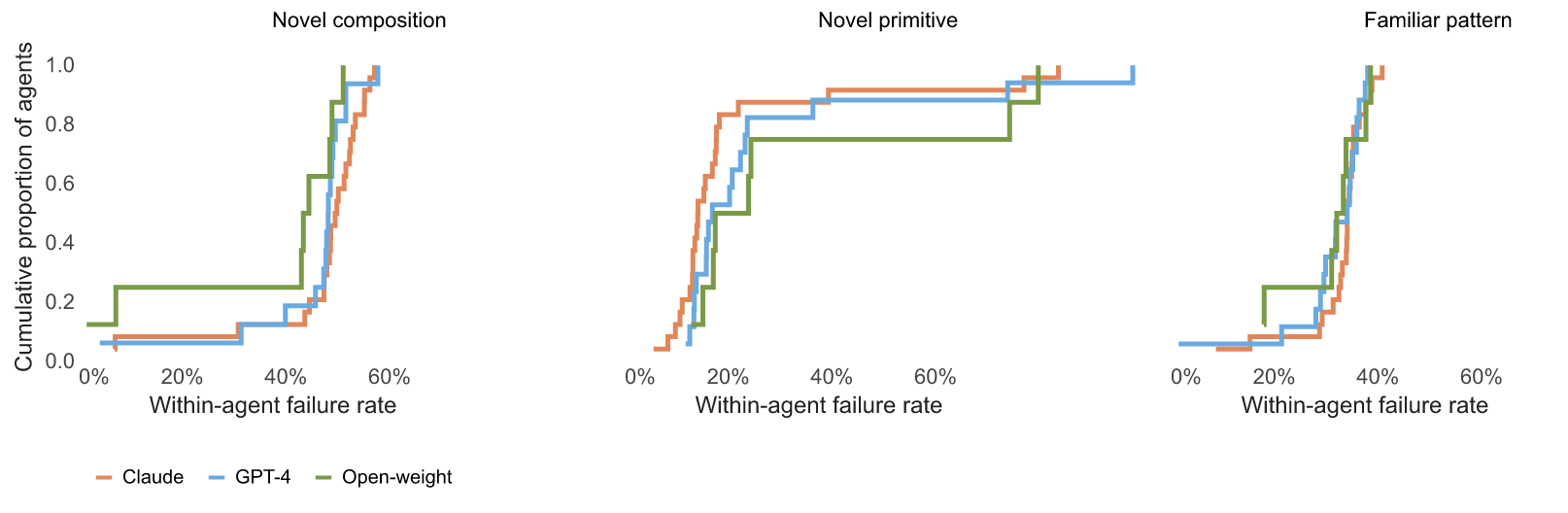}
      \caption{Failure type distributions by model family.}
      \label{fig:failure_by_family}
  \end{subfigure}
  \caption{Model behavior on compositional tasks.}
  \label{fig:composition_analysis}
  \end{figure}

  With this data, we can also validate other kinds of inferences regarding the shapes of problems and model behavior. For
  example, past work has found that models struggle with compositional generalization---this means that they are brittle in
  contexts where they need to do incremental work. We validate this in the wild and find that problems that require
  compositional generalization, defined by problems that have nested procedures, are especially difficult for agents, as
  evidenced by their failure rates. We define compositional problems as ones where the subproblems are ones that have been
  solved by the agent before, but when put together the agent failed. Looking at Figure~\ref{fig:compositional_failure} we
  can observe that compositionality is a relatively strong signal of failure, followed by novel actions, and familiar
  procedures yield the lowest failure rates. However, the pattern holds the most strongly for RLHF-trained models and much
  more modest for more classically trained models. Across models, similar trends hold, but open-weight agents have lower
  failure rates to compositional problems---when computing the difference of open-weight model failure at all points
  compared to other model families is ${\sim}9\%$.

  Next, developers make a number of decisions to train and scaffold their models. We set out to understand how this affects
  model processes based on a set of tasks and find that ``fingerprints'' are recoverable---unique model-specific patterns
  regarding tool-use and order of operations that can be attributed back to an agent. Given certain parameters like the
  costs for different types of actions and the length of a task, hopefully this can form the basis of more informed model
  choice and configuration decisions depending on a developer's goal (i.e.\ task and workflow-aware routing). For a given
  model, its procedural space can be described by its entropy and compression, where entropy describes the variance of the
  actions and compression is a rough proxy for repetitiveness. We find the GPT-4o has the most diversity of actions
  compared to Claude models.

  Entropy is just one method of characterizing procedural behavior given a trajectory and it makes sense that there is some
  level of behavioral similarity for models of the same family and harness---suggesting that architectural design choices
  bleed into procedural tendencies. We compare the divergence of procedural distributions with JSD to test and find that
  scaffolds and model generation are the strongest indicators of procedural habits compared to the model generation in distilled model
  pairs, suggesting that the generation of the model and programmed behavior play a disproportionate role and that a distilled student's
  procedural style is largely derived from its teacher. See Figure~\ref{fig:jsd_full} for a matrix showing all the pairwise divergence rates across
  models.

  \begin{table}[H]
  \centering\small
  \begin{tabular}{llrr}
  \toprule
  \textbf{Factor} & \textbf{Comparison} & \textbf{JSD} & \textbf{improvement from baseline} \\
  \midrule
  Lineage  & teacher $\to$ distilled child       & 0.250 & 3.0 \\
  Family      & within family, across generations   & 0.518 & 6.3 \\
  Scaffold & same model, two harnesses           & 0.533 & 6.5 \\
  \bottomrule
  \end{tabular}
  \caption{Procedural divergence across agent distributions when grouped by agent type.}
  \end{table}

  \subsection{How agents acquire context}
  We also look at the breadth of files that agents access. We believe this is a measure of efficiency and localization
  ability. In other words, the file coverage is a proxy for the range of the codebase that the agent interacts with. Claude-3.5 has the largest
  range, with a mean of 2.41 files per task, compared to an average of 1.66-1.90. However, the pass
  rate of the model is not proportional to the number of files that it accesses, where for every model there is usually a
  tipping point where accessing files becomes a signal of struggle rather than success. We observe this more broadly across agents where the
  highest pass rate instances are correlated with fewer files being edited, while there is a drop in pass rate (steeper for older models and
  milder for newer ones) when file edits increase---likely suggesting that success on a problem is tied to how easy it is to pinpoint.

  \begin{figure}[H]
      \centering
      \includegraphics[width=0.6\linewidth]{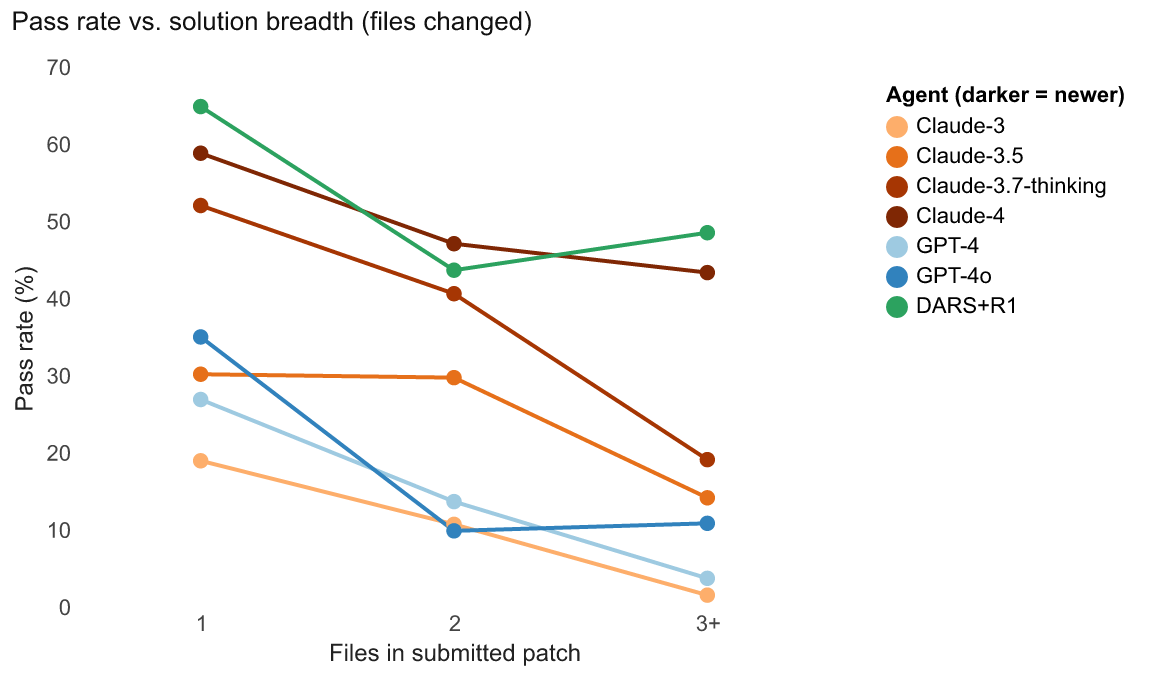}
      \caption{Figure plotting agent pass rate by the number of files changed. We observe that a low file change rate is lightly correlated with
  pass rate, potentially because it means the scope of the solution was more straightforward, compared to complex tasks that are either hard to
  localize or require a breadth of files as context to solve. }
      \label{fig:files_by_pass_rate}
  \end{figure}

  % \textcolor{purple}{TODO: Add multi-file exploration findings here as proper prose --- GPT-4: 23.5\%→7.7\% at 3+ files,
  % Claude-3.5 least degradation (29.0\%→17.6\%), mean 2.41 files opened vs 1.66-1.90 for peers. Source: rich\_features
  %  analysis.}

  We can also use our framework to understand trajectory anomalies as it pertains to context retrieval to offer developers useful signal for
  when their agents
  are failing. For instance, out of all 63 times that Agentless ``wins''---where it successfully passes a task that
  SWE-agent does not---SWE-agent's failures are marked by long edit streaks: it is over twice as likely to make five or
  more edits in a row when it fails as when it succeeds. Agentless is an overall more efficient agent, as it will require
  an average of 13 steps, where it will localize the error, patch the problem, and validate in a clean loop, compared to
  SWE-agent's average of 33.

  \begin{table}[H]
  \centering\small
  \begin{tabular}{lrrrr}
  \toprule
  \textbf{Outcome} & \textbf{$n$} & \textbf{Difficulty} & \textbf{SWE-agent steps} & \textbf{Agentless steps} \\
  \midrule
  Both pass       &  54 & 7.17 & 20.4 & 13.0 \\
  SWE-agent only  &  15 & 4.53 & 30.7 & 13.0 \\
  Agentless only  &  63 & 4.59 & 33.3 & 13.0 \\
  Neither passes  & 157 & 0.69 & 36.8 & 13.0 \\
  \midrule
  \multicolumn{2}{l}{Total} & & 289 \\
  \bottomrule
  \end{tabular}
  \caption{Outcomes of two agent scaffolds on 289 tasks.}
  \label{tab:scaffold_outcomes}
  \end{table}

  \section{Controlled evaluations with procedural controls}
  In agent programming and benchmarking, an engineer can specify parameters such as the number of tool use turns and
  response lengths in the form of tokens, but there is little control over how a task is done. Controlling for behavior at
  the level of actions is an under-explored way of assessing capabilities---while ablations are possible at different
  levels of abstraction, it is not possible to say a model can only retrieve context via grep or when, or run a certain
  number of tests and after what actions.

  Fingerprints give us behavioral trends that form the basis for predicting which actions a given agent is most likely to
  take, useful for next-action prediction and anomaly detection. We batch trajectories into five groups, training models on
  four and leaving one as a held-out test set using GroupKFold and see that fingerprinting accuracy is relatively
  unchanged when grouping by task ($\Delta = -0.007$), demonstrating that identifiability is largely a function of style
  and not task memorization from the probe itself (for example, pattern matching certain actions to specific tasks).

  As shown in Table~\ref{tab:trajectory_holdouts}, agents fall into three regimes: deterministic scaffolds like Agentless,
  which are nearly fully predictable (82\% accuracy, $+$51 points over baseline); constrained scaffolds like Moatless,
  where the scaffold drives behavior and the baseline is already high (73\% accuracy, $+$8 points); and open-ended
  model-driven agents like Claude and GPT, where predictions are stronger at the single-action level (e.g., Claude-3.5 at
  50\% accuracy, $+$33 points). We also find
  that edit streaks are a strong failure signal: for Moatless+DeepSeek-V3, 43\% of trajectories have edit streaks of 5 or
  more, and of those, the fail rate is 80\%, compared to 59\% for no-streak trajectories. For Claude-3.5 the signal is
  weaker, with pass rate dropping from 26\% to 11\% with a streak $\geq$5, though the magnitude is comparable. Overall, we believe this makes a
  strong case for procedural representations as a basis for next-action and final state prediction.

  \begin{table}[htbp!]
  \centering\small\setlength{\tabcolsep}{4pt}
  \begin{tabular}{lrrrrrr}
  \toprule
  & \multicolumn{3}{c}{\textbf{Next action}} & \multicolumn{3}{c}{\textbf{Next stage}} \\
  \cmidrule(lr){2-4}\cmidrule(lr){5-7}
  \textbf{Agent} & Base & Acc. & $\Delta$ & Base & Acc. & $\Delta$ \\
  \midrule
  \rowcolor{claudebg} Claude-3              & 13\% & 46\% & $+$33 & 33\% & 47\% & $+$14 \\
  \rowcolor{claudebg} Claude-3.5            & 18\% & 50\% & $+$33 & 35\% & 51\% & $+$16 \\
  \rowcolor{claudebg} Claude-3.7-thinking   & 23\% & 54\% & $+$31 & 44\% & 65\% & $+$21 \\
  \rowcolor{claudebg} Claude-4              & 51\% & 61\% & $+$10 & 51\% & 67\% & $+$16 \\
  \midrule
  \rowcolor{gptbg} GPT-4                 & 20\% & 57\% & $+$37 & 50\% & 56\% & $+$6  \\
  \rowcolor{gptbg} GPT-4o                & 27\% & 60\% & $+$34 & 50\% & 62\% & $+$12 \\
  \midrule
  \rowcolor{owbg} DARS+DeepSeek-R1      & 30\% & 46\% & $+$17 & 37\% & 49\% & $+$12 \\
  \rowcolor{owbg} Agentless+Claude-3.5  & 31\% & 82\% & $+$51 & 31\% & 82\% & $+$51 \\
  \rowcolor{owbg} Moatless+DeepSeek-V3  & 65\% & 73\% & \textbf{$+$8} & 65\% & 73\% & \textbf{$+$8} \\
  \bottomrule
  \end{tabular}
  \caption{Next-action and next-stage prediction accuracy across agent trajectories.}
  \label{tab:trajectory_holdouts}
  \end{table}

  \subsection{Programming procedural rewards}
  One application of \texttt{ProcGrep} is the ability to specify and reward model behavior on
  long-horizon tasks. Typically, reward systems for coding models have focused on binary outcomes, but trajectory
  specifications enable partial rewards based on procedural milestones, providing a denser signal that captures
  intermediate progress and can be more readily adapted to changing goals. With ProcGrep, rewards can be assigned to action
  sequences in uniquely fine-grained ways. Here, we design a specification to reward what we define as problem-solving best practices such as
  exploration, implementation, and test verification and to penalize long edit streaks that would burn resources and a lack of search which
  suggests unsystematic problem-solving.

  \begin{tcolorbox}[colback=black!3, colframe=black!30, boxrule=0.4pt, left=10pt, right=10pt, top=8pt, bottom=8pt,
  fontupper=\small]
  \texttt{phases:}\\
  \texttt{~~- name: exploration~~~~reward: 0.10}\\
  \texttt{~~~~require\_any: [\{atom: search\_repo\}, \{atom: read\_file\}]}\\
  \texttt{~~~~min\_occurrences: 2~~~~before\_first: edit}\\
  \texttt{~~- name: implementation~~~~reward: 0.15}\\
  \texttt{~~~~require\_any: [\{atom: edit\}]}\\
  \texttt{~~- name: test\_verification~~~~reward: 0.25}\\
  \texttt{~~~~require\_sequence: [edit, run\_test]~~~~max\_gap: 5}\\
  \texttt{~~- name: completion~~~~reward: 0.10}\\
  \texttt{~~~~require\_any: [\{atom: submit\}]}\\
  \texttt{penalties:}\\
  \texttt{~~- name: edit\_streak~~~~penalty: 0.15}\\
  \texttt{~~~~pattern: [edit, edit, edit, edit, edit]~~~~contiguous: true}\\
  \texttt{~~- name: no\_search~~~~penalty: 0.05}\\
  \texttt{~~~~require\_absent\_before: [search\_repo, read\_file]~~~~before\_first: edit}\\
  \texttt{bonuses:}\\
  \texttt{~~- name: test\_driven~~~~reward: 0.10}\\
  \texttt{~~~~require\_sequence: [run\_test]~~~~before\_first: edit}
  \end{tcolorbox}

  We show a snippet of the spec in YAML above and provide the point scheme below and a table of results where we show that Agentless+Claude-3.5
  is the winning scaffold for following said instructions. Here, we hand-author an action sequence and the feedback derived from it, but do not
  close the loop with a trained model. Thus, we use the following as an illustrative example and recognize that crafting a state will depend on
  a variety of goals and decisions:

  \begin{tcolorbox}[colback=black!3, colframe=black!30, boxrule=0.4pt, left=10pt, right=10pt,
  top=8pt, bottom=8pt, fontupper=\small]
  \texttt{\{"instance\_id": "django\_\_django-12345",}\\
  \texttt{~~"binary\_pass": false,~~~~"proc\_score": 0.35,}\\
  \texttt{~~"satisfied\_phases": ["exploration",}\\
  \texttt{~~~~~~~~~~~~~~~~~"implementation", "test\_verification"],}\\
  \texttt{~~"triggered\_penalties": ["edit\_streak"],}\\
  \texttt{~~"triggered\_bonuses": []\}}
  \end{tcolorbox}

  \begin{table}[H]
  \centering\small
  \begin{tabular}{lllr}
  \toprule
  \textbf{Type} & \textbf{Component} & \textbf{Triggered when} & \textbf{Points} \\
  \midrule
  Phase   & exploration        & $\geq 2$ search/read before first edit & $+0.10$ \\
          & implementation     & any edit                               & $+0.15$ \\
          & test\_verification & edit $\to$ run\_test within 5 steps     & $+0.25$ \\
          & completion         & any submit                             & $+0.10$ \\
  \midrule
  Bonus   & test\_driven       & run\_test before first edit            & $+0.10$ \\
  \midrule
  Penalty & edit\_streak       & 5 contiguous edits                     & $-0.15$ \\
          & no\_search         & first edit with no prior search/read   & $-0.05$ \\
  \bottomrule
  \end{tabular}
  \caption{Point scheme for the procedural reward spec. Scores are bounded to the $[0,1]$ range and the maximum attainable score is $0.70$).}
  \label{tab:reward_scheme}
  \end{table}

  \begin{table}[H]
  \centering\small\setlength{\tabcolsep}{5pt}
  \begin{tabular}{lrrrrrrr}
  \toprule
                      &     & \textbf{Proc.} & \textbf{Explore} & \textbf{Test} & \textbf{Test-} & \textbf{Edit} & \textbf{No} \\
  \textbf{Agent}      & \textbf{$n$} & \textbf{score} &        & \textbf{verif.} & \textbf{driven} & \textbf{streak} & \textbf{search} \\
  \midrule
  \rowcolor{owbg}     Agentless+Claude-3.5 & 300 & 0.600 & 100\% & 100\% & 0\%  & 0\%  & 0\%  \\
  \rowcolor{owbg}     DARS+R1              & 300 & 0.468 & 33\%  & 86\%  & 0\%  & 8\%  & 34\% \\
  \rowcolor{claudebg} Claude-3.5           & 289 & 0.447 & 36\%  & 90\%  & 1\%  & 16\% & 54\% \\
  \rowcolor{claudebg} Claude-3.7-thinking  & 284 & 0.397 & 97\%  & 18\%  & 17\% & 2\%  & 0\%  \\
  \rowcolor{claudebg} Claude-3             & 300 & 0.384 & 19\%  & 88\%  & 9\%  & 12\% & 73\% \\
  \rowcolor{gptbg}    GPT-4o               & 278 & 0.383 & 23\%  & 97\%  & 2\%  & 32\% & 70\% \\
  \rowcolor{gptbg}    GPT-4                & 300 & 0.381 & 21\%  & 85\%  & 0\%  & 21\% & 75\% \\
  \rowcolor{claudebg} Claude-4             & 288 & 0.345 & 100\% & 0\%   & 1\%  & 3\%  & 0\%  \\
  \rowcolor{owbg}     Moatless+V3          & 300 & 0.187 & 49\%  & 0\%   & 0\%  & 41\% & 0\%  \\
  \bottomrule
  \end{tabular}
  \caption{Procedural reward breakdown.}
  \label{tab:reward_results}
  \end{table}

  % \textcolor{blue}{Scoring the distilled child model yields mean procedural score 0.795 across 498 trajectories---with
  %passing trajectories scoring 0.902 versus 0.723 for failing ones ($\Delta = 0.179$). Notably, 87.8\% of child
  % trajectories trigger the \texttt{stuck\_reading} penalty, compared to 87.7\% of the teacher---confirming this failure
  % mode was inherited, not introduced by fine-tuning. The teacher achieves mean procedural score 0.940 with a tighter
  % pass/fail gap (0.953 vs.\ 0.927). The full reward spec and scorer are available at \texttt{from procgrep.reward import
  % load\_spec, score}.}

  In the following, we do a run to compare two procedural types (test-driven and patch-driven) across agent rollouts. For the test-driven
  specification, we reward interleaving edits with tests; for the patch-driven one, we reward the model going directly from edits to submission
  and penalize tests in this regime. With this, we can interpret the $\Delta$ as how consistent each regime is with the model's innate
  procedural tendencies (Table~\ref{tab:spec_comparison}).

  \begin{table}[H]
  \centering\small
  \begin{tabular}{lrrr}
  \toprule
  \textbf{Agent} & \textbf{Test-driven} & \textbf{Patch-driven} & \textbf{$\Delta$} \\
  \midrule
  \rowcolor{gptbg}    GPT-4o                & 0.562 & 0.399 & $+0.163$ \\
  \rowcolor{claudebg} Claude-3              & 0.541 & 0.399 & $+0.143$ \\
  \rowcolor{gptbg}    GPT-4                 & 0.515 & 0.452 & $+0.063$ \\
  \rowcolor{claudebg} Claude-3.5            & 0.565 & 0.513 & $+0.052$ \\
  \rowcolor{owbg}     Agentless+Claude-3.5  & 0.700 & 0.700 & $0.000$ \\
  \rowcolor{owbg}     DARS+R1               & 0.550 & 0.557 & $-0.007$ \\
  \rowcolor{owbg}     Moatless+V3           & 0.095 & 0.456 & $-0.361$ \\
  \rowcolor{claudebg} Claude-3.7-thinking   & 0.285 & 0.677 & $-0.391$ \\
  \rowcolor{claudebg} Claude-4              & 0.200 & 0.692 & $-0.492$ \\
  \bottomrule
  \end{tabular}
  \caption{Mean procedural score under a test-driven vs.\ patch-driven spec, where $\Delta>0$ shows that a test-driven regime is favored.}
  \label{tab:spec_comparison}
  \end{table}

  With these examples, we hope to inspire what is possible to study, though we believe there are a number of configurations one could program
  depending on the use case. The scores we assign to the rewards are relatively arbitrary---our main contribution is building the means to
  specify them in the first place---and a meaningful area of future work is dynamic scoring that adapts to observed final states in
  continual-learning contexts.

  \section{Do students behave like their teachers? (a distilled model case study)}
  \label{sec:distillation}
  We consider a distilled model pair consisting of a teacher and a student model, where the student model is trained on trajectories generated
  by the teacher through supervised fine-tuning. While distillation is typically employed because it can produce performant models at a fraction
  of the cost, do students actually learn to problem solve like their parents and what value does this confer if they do
  \citep{hinton2015distillingknowledgeneuralnetwork, zhang2026embarrassinglysimpleselfdistillationimproves}? We find that the student inherits
  the teacher's full action vocabulary and shows concentrated entropy, which means that when the distilled model succeeds it has lower entropy
  and is more procedurally similar than when it fails, showing that procedural mimicry could be a means to improve success rates for distilled
  models. We analyze 284 parent and 498 child trajectories using \texttt{lineage\_diff} across three axes: vocabulary preservation, entropy
  shifts, and outcome divergence. Stratifying by outcome at the native level, passing child trajectories share 0.204 of the parent's tool
  signatures, whereas failing trajectories share only 0.193 ($\Delta = 0.01$).

  %SWE-agent-LM-32B is produced by the SWE-smith pipeline \citep{yang2025swesmithscalingdatasoftware}: Claude-3.7 Sonnet (the teacher) generates
 % trajectories on SWE-bench tasks, passing trajectories are filtered and used as supervised demonstrations, and Qwen2.5-32B is fine-tuned on the
 % resulting dataset. Supervision is on \emph{outputs only}---the student learns to reproduce the teacher's actions but not the reasoning behind
%  them.

  \begin{figure}[H]
  \centering
  \begin{subfigure}[t]{0.46\linewidth}
      \includegraphics[width=\linewidth]{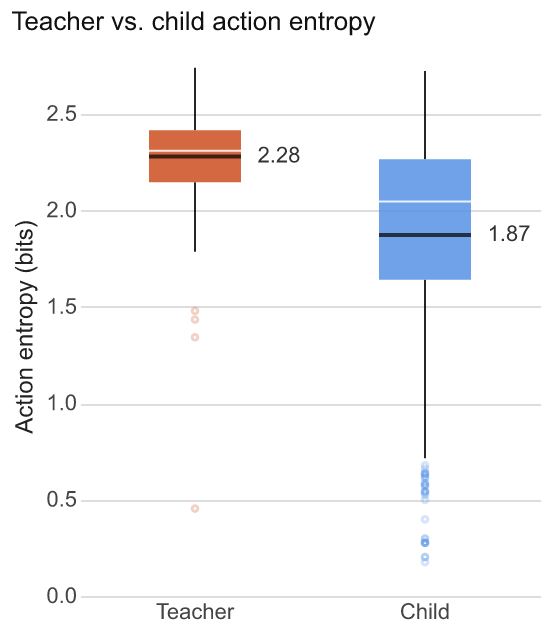}
      \caption{Action entropy student and teacher model pair where the student's action distribution is more
  concentrated  ($-0.41$ bits).}
      \label{fig:dist_entropy}
  \end{subfigure}\hfill
  \caption{Comparing the trajectories of distilled student-teacher pairs, comparing Claude-3.7~Sonnet (teacher) $\to$ SWE-agent-LM-32B (child)
  on SWE-bench Verified.}
  \label{fig:distillation}
  \end{figure}

  % \begin{subfigure}[t]{0.46\linewidth}
    %  \includegraphics[width=\linewidth]{figures/fig_distillation_B_conditional.png}
   %   \caption{\textcolor{blue}{Conditional JSD by prefix action (canonical). Divergence is modest and spread, not
  % localized to one control point.}}
  %    \label{fig:dist_conditional}
  % \end{subfigure}

  \section{Conclusion}

  In this work we set out to define and motivate a new lens for studying agent trajectories, specifically in coding settings. They are natural
  exhaust from models that are being deployed for increasingly long-horizon tasks, and we believe there is a need for tools to make use of them
  to help a new age of developers who are tasked with deciding which models to employ and how to architect their agents. We validate procedural
  fingerprinting, demonstrating that problem-solving style is a real property of agent behavior and something that can be programmed. We find
  that agents are recoverable from their traces alone at a near 90\% rate as compared to random baseline of 11\%. We show that different models
  that would otherwise be studied in isolation can be compared through new methods for representing their traces and breaking them into
  meaningful sub-processes. We also show that certain procedural quirks are linked to success and failure, which can be the basis of inferring
  outcomes even when a full trajectory has not been carried out in its entirety, suggesting that telemetry can be a way to reduce overall
  compute spend.

  We are excited about new ways of applying ProcGrep--the library that was built to do the top-down analysis we put forth in this
  work---particularly, when training models to carry out long-horizon tasks with resource management in mind. In the future, we anticipate the
  methods put forth in this work can complement model diffing methods built to probe into models at a functional and behavioral level to produce
  more holistic analyses of agents \citep{jiralerspong2026crossarchitecturemodeldiffingcrosscoders, aranguri2025modeldiff}.

  %\textcolor{blue}{Prior work on behavioral diffing and model-diffing has compared model outputs at the response level.
   % Three directions follow from this work. First, \emph{lineage attribution at scale}: applying \texttt{lineage\_diff}
   % across pairs of open-weight models on HuggingFace to ask whether procedural fingerprints can identify training
   % relationships when the relationship is not disclosed. Second, \emph{procedural regression detection} across model
   % releases---the Claude-3.7$\to$Claude-4 opening-move shift (JSD\,=\,0.27 at step~1) shows that training-recipe changes
  %  surface as procedural discontinuities. Third, correlating procedural axes with model attributes through observational
   %regression; this can generate hypotheses for controlled distillation experiments like the one in
   % \S\ref{sec:distillation}.}

  \bibliographystyle{plainnat}
  \bibliography{main}

  \appendix

  \section{Appendix}

  \subsection{Algorithms for vocabulary discovery}

  \subsubsection{Byte-Pair Encoding}

  \begin{algorithm}[H]
  \caption{BPE vocabulary induction}
  \begin{algorithmic}[1]
  \Require action sequences $S=\{s_1,\dots,s_n\}$; target size $K$
  \Ensure vocabulary $V$; tokenized sequences $T$
  \State $V \gets \{\text{all atoms appearing in } S\}$
  \State $T \gets S$ \Comment{each action starts as its own token}
  \While{$|V| < K$}
    \State count frequency of every \emph{adjacent} token pair $(a,b)$ in $T$
    \State $(a^\star,b^\star) \gets \arg\max_{(a,b)} \mathrm{freq}(a,b)$
    \If{$\mathrm{freq}(a^\star,b^\star) = 0$} \textbf{break} \EndIf
    \State $t \gets a^\star\!+\!b^\star$;\quad $V \gets V \cup \{t\}$
    \State replace every adjacent occurrence of $(a^\star,b^\star)$ in $T$ with $t$
  \EndWhile
  \State \Return $V,\;T$
  \end{algorithmic}
  \end{algorithm}

  \subsubsection{PrefixSpan}

  \begin{algorithm}[H]
  \caption{PrefixSpan (single-item event sequences)}
  \begin{algorithmic}[1]
  \Require sequences $S$; minimum support $\sigma$
  \Ensure frequent sequential patterns $P$
  \Function{PrefixSpan}{$\alpha,\; S|_\alpha$}
    \For{each action $b$ with $\mathrm{support}(b,\, S|_\alpha) \ge \sigma$}
      \State $\alpha' \gets \alpha \cdot b$;\quad $P \gets P \cup \{\alpha'\}$
      \State $S|_{\alpha'} \gets \{\, \text{suffix of } r \text{ after its first } b : r \in S|_\alpha \,\}$
      \State \Call{PrefixSpan}{$\alpha',\; S|_{\alpha'}$}
    \EndFor
  \EndFunction
  \State $P \gets \emptyset$;\quad \Call{PrefixSpan}{$\langle\rangle,\; S$} \Comment{$S|_{\langle\rangle}=S$}
  \State \Return $P$
  \end{algorithmic}
  \end{algorithm}

   \begin{table}[t]
  \centering
  \caption{Agreement and compositionality metrics across LLM judges.}
  \label{tab:judge_comparison}
  \begin{tabular}{lccc}
  \hline
  \textbf{Judge} & \textbf{\% Compositional} & \textbf{F1 (k=1) Pass/Fail} & \textbf{Mean $\kappa$} \\
  \hline
  GPT-4o          & 38.1\% & 0.363 & 0.136 \\
  GPT-4o-mini     & 36.5\% & 0.246 & 0.156 \\
  Qwen-2.5-72B    & 17.2\% & 0.167 & 0.128 \\
  Llama-3.3-70B   & 21.8\% & 0.000 & 0.123 \\
  \hline
  \end{tabular}
  \end{table}
  \subsection{Classifier prompt}
   \begin{tcolorbox}[enhanced,
    colback=black!3, colframe=black!30, boxrule=0.5pt, arc=2pt,
    left=10pt, right=10pt, top=8pt, bottom=8pt, fontupper=\small,
    title=\textbf{Judge classifier prompt},
    coltitle=black!75, colbacktitle=black!8, fonttitle=\small,
    attach boxed title to top left={xshift=6pt, yshift=-3pt},
    boxed title style={colframe=black!30, arc=1pt}]

  You are evaluating a software engineer's response to a bug report. You are given the
  reference patch---the actual fix---and use it as ground truth.

  \medskip
  \begin{tabular}{@{}ll@{}}
  \texttt{Bug report:}      & \texttt{\{problem\_statement\}} \\
  \texttt{Reference patch:} & \texttt{\{gold\_patch\}} \\
  \texttt{Response:}        & \texttt{\{response\}} \\
  \end{tabular}

  \medskip
  Reason briefly about each dimension, then score it $0$--$3$ ($0=$ wrong, $3=$ exact
  match to the reference):
  \begin{itemize}\setlength\itemsep{1pt}
    \item \emph{localization} --- same file/function as the patch?
    \item \emph{edit\_type} --- right kind of change?
    \item \emph{plan\_quality} --- would this plan reproduce the patch?
    \item \emph{explanation} --- grounded in what the patch does?
  \end{itemize}

  Identify the first plan step where the response diverges ($-1$ if fully aligned) and its level:
  \begin{itemize}\setlength\itemsep{1pt}
    \item \emph{surface} --- wrong file/function name (token-level)
    \item \emph{compositional} --- right location, wrong operation or order (syntactic-level)
    \item \emph{relational} --- right operations, wrong dependency or interaction (graph-level)
    \item \emph{none} --- no divergence
  \end{itemize}

  \tcblower
  Output per-dimension reasoning, then:

  {\ttfamily\small
  \{"localization": int, "edit\_type": int, "plan\_quality": int,\\
  \phantom{\{}"explanation": int, "first\_deviation\_step": int,\\
  \phantom{\{}"divergence\_level": "surface|compositional|relational|none"\}}

  \end{tcolorbox}

      \begin{table}[h]
  \centering\small
  \begin{tabular}{lcccc}
  \toprule
  \textbf{Judge model} & \textbf{n} & \textbf{F1 (k=1)} & \textbf{\% compositional} & \textbf{Mean $\kappa$} \\
  \midrule
  GPT-4o         & 286 & 0.363 & 38.1\% & 0.136 \\
  GPT-4o mini    & 285 & 0.246 & 36.5\% & 0.156 \\
  Qwen 2.5 72B   &  87 & 0.167 & 17.2\% & 0.128 \\
  Llama 3.3 70B  &  87 & 0.000 & 21.8\% & 0.123 \\
  \midrule
  Weighted mean  & 745 & 0.253 & --     & --     \\
  \bottomrule
  \end{tabular}
  \caption{Comparing judges and their agreement. Overall, agreement is low across models with mean Cohen scores ($\kappa$) of ~0.12–0.16. GPT
  family models have low agreement rates with open-weight models and same-family model pairs and open-weight models tend to agree with each
  other.}
  \label{tab:divergence_by_model}
  \end{table}

    \subsection{Procedural reward specifications}
  The test-driven and patch-driven specifications scored in Table~\ref{tab:spec_comparison}. Both have a maximum attainable
  score of $0.70$ before penalties.
  \\
  \\
  \begin{tcolorbox}[
    colback=black!3,
    colframe=black!30,
    boxrule=0.4pt,
    left=10pt,
    right=10pt,
    top=8pt,
    bottom=8pt,
    fontupper=\small,
    title=\texttt{test\_driven},
    coltitle=black!75,
    colbacktitle=black!8,
    fonttitle=\small,
    before skip=0pt,
    after skip=0pt
  ]
  \texttt{phases:}\\
  \texttt{~~- name: exploration~~~~reward: 0.10}\\
  \texttt{~~~~require\_any: [\{atom: search\_repo\}, \{atom: read\_file\}]}\\
  \texttt{~~~~min\_occurrences: 2~~~~before\_first: edit}\\
  \texttt{~~- name: test\_first~~~~reward: 0.20}\\
  \texttt{~~~~require\_sequence: [run\_test]~~~~before\_first: edit}\\
  \texttt{~~- name: implementation~~~~reward: 0.10}\\
  \texttt{~~~~require\_any: [\{atom: edit\}]}\\
  \texttt{~~- name: verification~~~~reward: 0.30}\\
  \texttt{~~~~require\_sequence: [edit, run\_test]~~~~max\_gap: 5}\\
  \texttt{penalties:}\\
  \texttt{~~- name: edit\_streak~~~~penalty: 0.15}\\
  \texttt{~~~~pattern: [edit, edit, edit, edit, edit]~~~~contiguous: true}
  \end{tcolorbox}%
  \begin{tcolorbox}[
    colback=black!3,
    colframe=black!30,
    boxrule=0.4pt,
    left=10pt,
    right=10pt,
    top=8pt,
    bottom=8pt,
    fontupper=\small,
    title=\texttt{patch\_driven},
    coltitle=black!75,
    colbacktitle=black!8,
    fonttitle=\small,
    before skip=0pt,
    after skip=0pt
  ]
  \texttt{phases:}\\
  \texttt{~~- name: localization~~~~reward: 0.20}\\
  \texttt{~~~~require\_any: [\{atom: search\_repo\}, \{atom: read\_file\}]}\\
  \texttt{~~~~min\_occurrences: 2~~~~before\_first: edit}\\
  \texttt{~~- name: implementation~~~~reward: 0.30}\\
  \texttt{~~~~require\_any: [\{atom: edit\}]}\\
  \texttt{~~- name: completion~~~~reward: 0.20}\\
  \texttt{~~~~require\_any: [\{atom: submit\}]}\\
  \texttt{penalties:}\\
  \texttt{~~- name: test\_thrash~~~~penalty: 0.15}\\
  \texttt{~~~~pattern: [run\_test, run\_test, run\_test]~~~~contiguous: true}\\
  \texttt{~~- name: edit\_streak~~~~penalty: 0.10}\\
  \texttt{~~~~pattern: [edit, edit, edit, edit, edit]~~~~contiguous: true}
  \end{tcolorbox}

  %\subsection{\textcolor{blue}{Serving and per-step capture}}
  %  \label{app:serving}
   % \textcolor{blue}{To expose per-step token counts for live procedural monitoring, serve the model with metrics and cache
  %  reporting enabled:}

  %  \begin{tcolorbox}[colback=black!3, colframe=black!30, boxrule=0.4pt, left=10pt, right=10pt, top=8pt, bottom=8pt,
   % fontupper=\small]
  %  \textcolor{blue}{
  %  \texttt{python -m sglang.launch\_server}~\textbackslash\\
   % \texttt{~~--model-path SWE-bench/SWE-agent-LM-32B}~\textbackslash\\
    %\texttt{~~--tp 1}~\textbackslash\\
    %\texttt{~~--enable-metrics}~\textbackslash\\
    %\texttt{~~--enable-cache-report}~\textbackslash\\
    %\texttt{~~--cuda-graph-max-bs 8}
    %}
    %\end{tcolorbox}

   % \textcolor{blue}{Each turn then returns \texttt{usage.completion\_tokens} and \texttt{cache\_hit\_tokens}; matched
    %against the \texttt{stuck\_reading} spec, these support a circuit-breaker that interrupts when the pattern completes
   % within 12 steps. These flags are the minimum for per-step capture; tensor-parallel and quantization settings are
   % model-dependent.}

  \begin{figure}[H]
      \centering
      \includegraphics[width=0.6\linewidth]{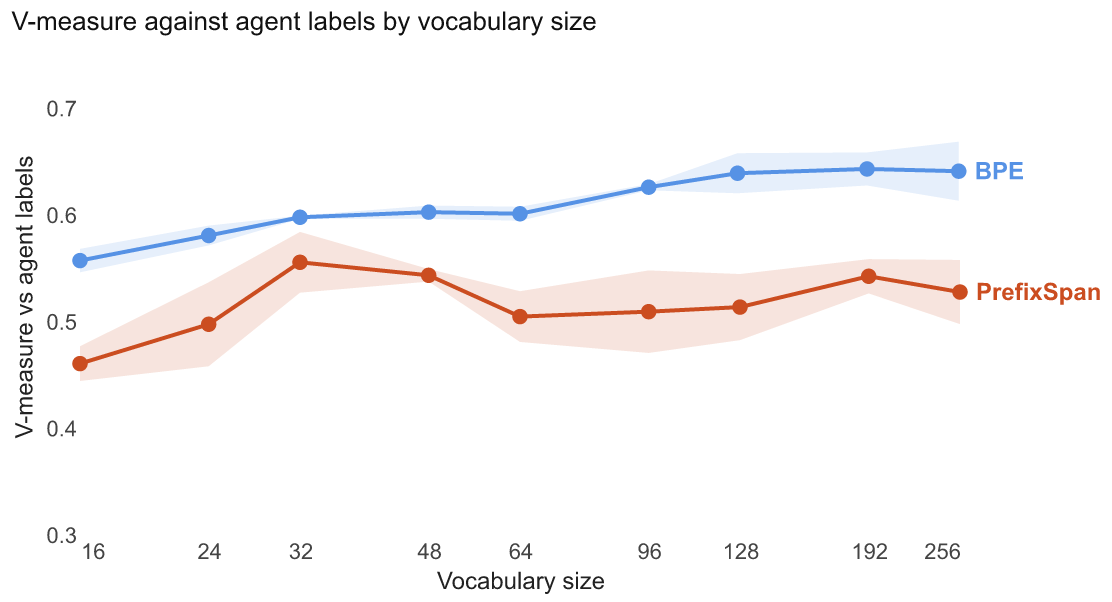}
   \caption{BPE separates agents better than PrefixSpan, scored by V-measure
  against agent labels. The golden set, derived from the evaluation dataset,
  treats a clustering as correct when every trajectory from a given agent is
  grouped with the others from that agent (e.g., all GPT-4o trajectories together,
  all Claude-3.5 trajectories together, and so on).}
      \label{fig:bpe_vs_prefixspan}
  \end{figure}

  \begin{figure}[H]
      \centering
      \includegraphics[width=0.62\linewidth]{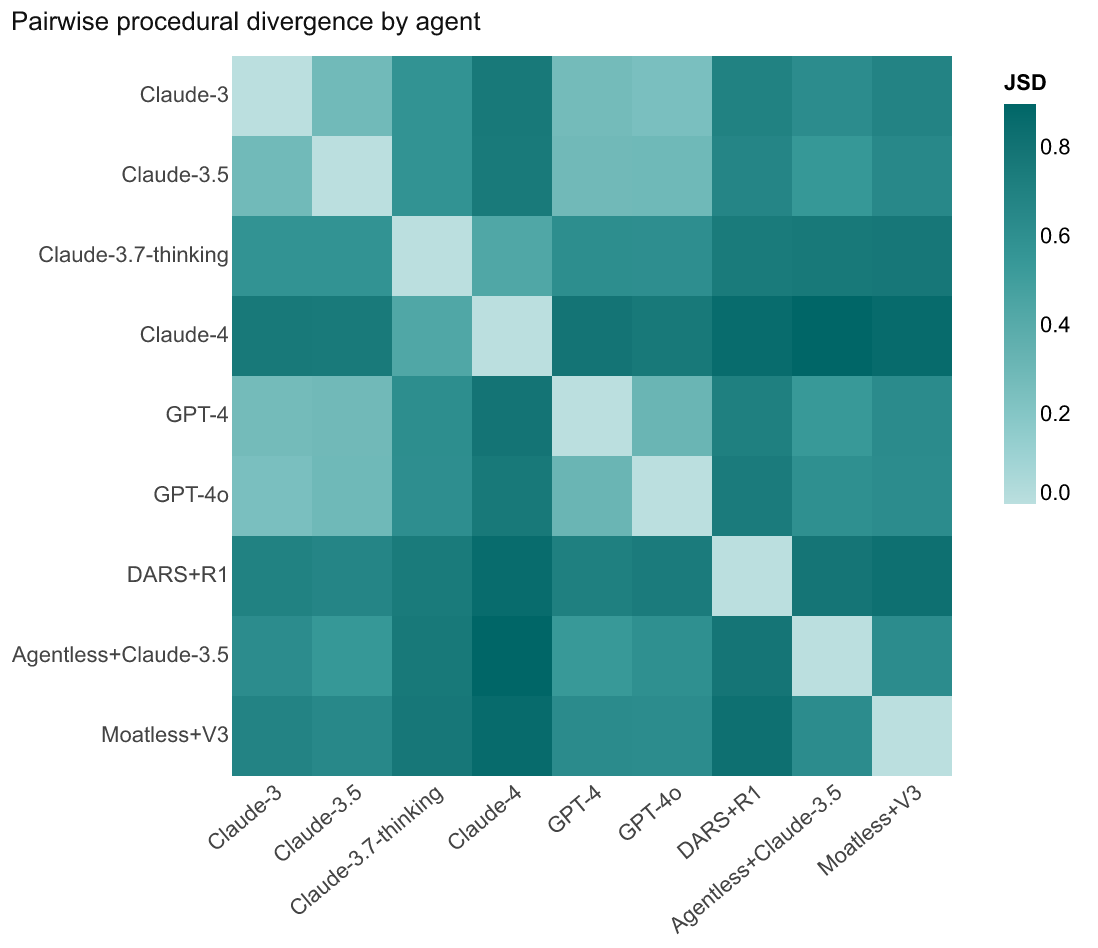}
   \caption{Ten agents compared by their pairwise divergence. Of all agent pairs, the distilled student (SWE-agent-LM-32B) is closest to its
  teacher (Claude-3.7~Sonnet), at JSD~$=~0.25$.}
      \label{fig:jsd_full}
  \end{figure}

  \begin{figure}[H]
      \centering
      \includegraphics[width=0.7\linewidth]{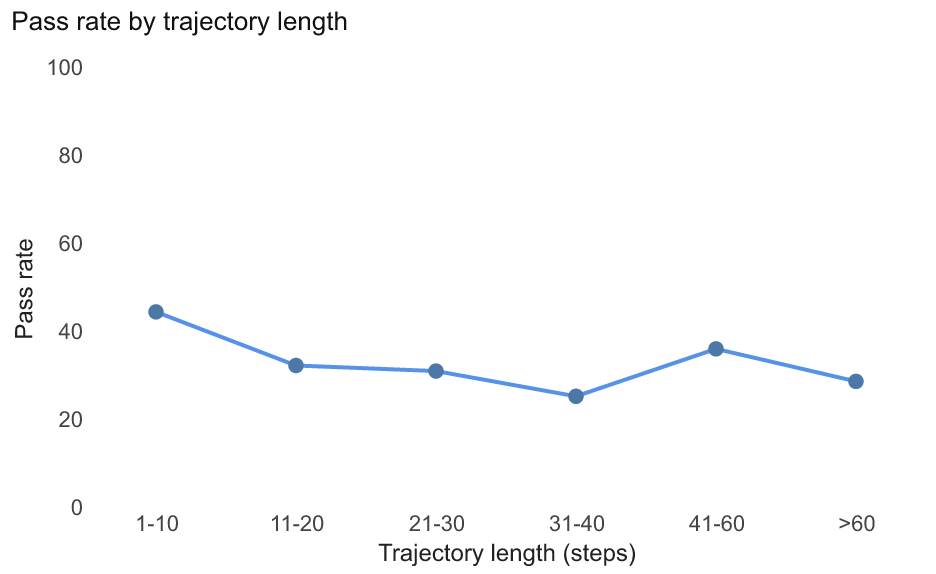}
      \caption{Pass rate by trajectory length (canonical-action count). Pass rate declines modestly and
  in a non-linear fashion, from 45\% at 1--10 steps to ${\sim}29\%$ beyond 60.}
      \label{fig:regression_length}
  \end{figure}

  \end{document}